\documentclass{emulateapj}

\usepackage{graphicx}
\usepackage{natbib}
\usepackage{apjfonts}
\usepackage{amsmath}

\bibpunct{(}{)}{;}{a}{}{,}

\begin{document}
\title{The Importance of Being First: Position Dependent Citation
  Rates on arXiv:astro-ph} 

\author{J.\,P. Dietrich} 
\affil{ESO, Karl-Schwarzschild-Str. 2, 85748 Garching b. M\"unchen,
  Germany} 
\email{jdietric@eso.org}

\begin{abstract}We study the dependence of citation counts of e-prints
  published on the arXiv:astro-ph server on their position in the
  daily astro-ph listing.
  Using the SPIRES literature database we reconstruct the astro-ph
  listings from July 2002 to December 2005 and determine citation
  counts for e-prints from their ADS entry. We use Zipf plots to
  analyze the citation distributions for each astro-ph position.
  We find that e-prints appearing at or near the top of the astro-ph
  mailings receive significantly more citations than those further
  down the list. This difference is significant at the $7\sigma$ level
  and on average amounts to two times more citations for papers at the
  top than those further down the listing. We propose three possible
  non-exclusive explanations for this positional citation effect and
  try to test them.
  We conclude that self-promotion by authors plays a role in the
  observed effect but cannot exclude that increased visibility at the
  top of the daily listings contributes to higher citation counts as
  well. We can rule out that the positional dependence of citations is
  caused by the coincidence of the submission deadline with the
  working hours of a geographically constrained set of intrinsically
  higher cited authors. We discuss several ways of mitigating the
  observed effect, including splitting astro-ph into several subject
  classes, randomizing the order of e-prints, and a novel approach to
  sorting entries by relevance to individual readers.
\end{abstract}

\section{Introduction}
\label{sec:introduction}

A number of studies looking at the influence of e-printing on citation
counts across disciplines \citep[e.g.,][]{2001Nature.411.521L} and in
Astronomy and Physics in particular
\citep[e.g.,][]{2004BAAS...36.1654S,2005BAAS...37..555M,2006JEPub...9....2H}
found that papers freely available online, particularly through the
arXiv e-print server\footnote{\texttt{http://www.arxiv.org/}}, are
cited more often than those not. This difference was studied in more
detail by \citet{2005IPM....41.1395K}, who proposed three
(non-exclusive) effects potentially responsible for higher citations
rates of journal articles also published as astro-ph e-prints. These
are defined as \citep[see][]{2005IPM....41.1395K}:
\begin{itemize}
\item The Open Access (OA) postulate -- Because the access to articles
  is unrestricted  by any payment mechanism authors are able to read
  them more easily, and thus they cite them more frequently;
\item The Early Access (EA) postulate -- Because the article appears
  sooner it gains both primacy and additional time in press, and is
  thus cited more;
\item The Self-selection Bias (SB) postulate -- Authors preferentially
  tend to promote (in this case by posting to the internet) the most
  important, and thus most citable articles.
\end{itemize}

\citet{2005IPM....41.1395K} found that open access to older issues of
astronomical journals through the Astrophysics Data System
\citep[ADS,][]{2000A&AS..143...41K} did not lead to increased citation
counts, plausibly denying a significant effect of the OA postulate.
The EA postulate was supported by a significant increase in citation
rate for recent papers. Concerning the SB postulate,
\citet{2005IPM....41.1395K} note that significantly fewer papers not
posted to astro-ph are among top 200 cited articles in the year
2003 Astrophysical Journal article than expected from the combined
OA+EA effect alone. It is thus established that EA and SB play a
significant role in gathering citations. 

We observed that on many days the arXiv astro-ph listing is headed by
one or more articles submitted a few seconds after the passing of the
deadline for submissions for the next astro-ph mailing. One example --
among many others -- is the astro-ph listing of Sep 24, 2007 headed by
three articles received within 20\,s after this list was started.
Considering that the number of new articles posted to astro-ph on a
typical day is about $35$, this temporal clustering close to the
submission deadline is conspicuous and suggests a peculiar form of
self-promotion. Apparently, a subset of authors expects higher
visibility and thus also higher citation rates for articles listed at
or close to the top of an astro-ph mailing.

In the following we study the dependence of citation counts on
articles' positions in the astro-ph listings. In Sect.~\ref{sec:data}
we describe how we reconstructed the daily astro-ph mailings for a
span of 2.5 years and count citations to e-prints. We analyze the
dependence of citations rates on astro-ph listing position in
Sect.~\ref{sec:analysis} and discuss three possible explanations for
the effect we find in our dataset. We present our conclusions in
Sect.~\ref{sec:disc-concl}.

\section{Data}
\label{sec:data}

Using the SPIRES High-Energy Physics Literature
Database\footnote{\texttt{http://www.slac.stanford.edu/spires/hep/}}
we reconstructed the daily astro-ph mailings in the period from July
2002 to December 2005. The starting date of our analysis is fixed by
the date from which on astro-ph is automatically added to SPIRES HEP
as soon as it is posted. Restricting SPIRES HEP queries by the date a
record was added allows us to request astro-ph listings for single
days. Older astro-ph articles were added to SPIRES HEP later in 2002.
They were easily recognized by their arXiv identifier and discarded
from our dataset. Remaining articles added out of order and occasional
small gaps were also found based on their arXiv identifiers. These
articles were associated with the correct posting date based on the
consecutive arXiv identifier if the association to a date was
unambiguous. Otherwise the article and the dates in question were
flagged as unreliable and excluded from the analysis. Numerous spot
checks confirmed that we were in this way able to reliably reproduce
daily astro-ph listings.

We chose not to include  e-prints appearing on astro-ph after December
2005 so that all articles had at least one year to gather citations
before we filled our citation  database in December 2006. Our database
contains astro-ph  listings for 839  days judged as reliable, with a
total of 27\,109 articles on these days.

We used NASA's Astrophysics Data System (ADS) Bibliographic
Services\footnote{\texttt{http://adsabs.harvard.edu/index.html}} to
determine if and where an e-print was published in printed form and to
determine the number of citations that an article has.
\citet{2006JEPub...9....2H} estimate that the concordance achieved by
the ADS between astro-ph e-prints and journal articles is 98\%. The
remaining incompleteness should not be a significant source of error,
especially considering that ADS also lists citations to arXiv
e-prints.

Astronomy is in the fortunate situation that only a few journals
dominate the publication of refereed articles. In agreement with
\citet{2005IPM....41.1395K} we consider these 7 core journals to be
\emph{The Astrophysical Journal (Letters)} and its \emph{Supplement
  Series}, \emph{Astronomy \& Astrophysics}, \emph{Monthly Notices of
  the Royal Astronomical Society}, \emph{The Astronomical Journal},
and \emph{Publications of the Astronomical Society of the Pacific}. We
set a flag in our database for those e-prints that are also published
in one of the core journals.

\section{Analysis}
\label{sec:analysis}

Random and systematical variations cause considerable fluctuations of
the length of the astro-ph listings. For our analysis we exclude
unusually short listings ($<15$ entries), which typically occur in the
period between Christmas and New Year, and extremely long postings
($>70$), of which we have 5 in our database, from our analysis. This
selection leaves 816 days to be considered.

\begin{figure}
  \plotone{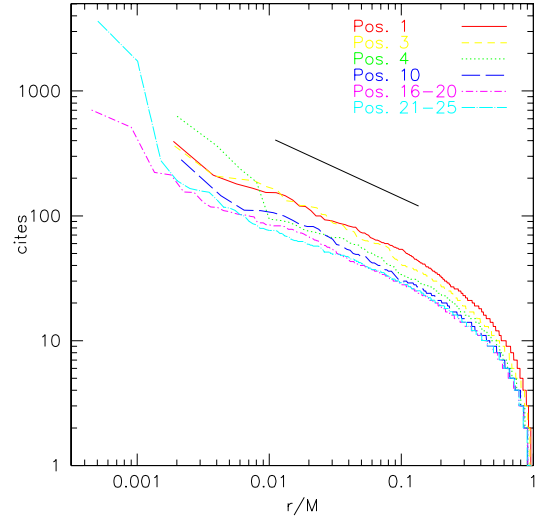}
  \caption{Zipf plot for different astro-ph positions. The $x$-axis
    shows the logarithm of the normalized rank of astro-ph postings
    after sorting them by citations. The $y$-axis shows the logarithm
    of the number of citations. The different line colors/styles
    encode the Zipf law for different astro-ph positions as given in
    the upper right corner of the figure. The solid black line
    indicates the slope of the power law. The high citation count of
    the top-ranked paper in the bin of astro-ph positions 21-25 is due
    to the WMAP first-year paper by \citet{2003ApJS..148..175S}, which
    has a submission time 1\,s before the deadline and appeared as
    23rd and last paper on Feb.~12, 2003. With 4118 citations reported
    by ADS at the time of writing, this is the most cited article in
    all of astronomy.}
  \label{fig:zipf}
\end{figure}
\citet{1998EPJB....4..131R} found that the citation distribution is a
power law over a large range of citation numbers. The mean and median
number of citations that papers of a given sample get are not good
estimators as the mean is strongly affected by a few highly cited
papers in the tail of the distribution while the median looks only at
the large number of very poorly cited papers. A better way to analyze
such data is in a Zipf plot. A Zipf plot shows the $r^\mathrm{th}$
most cited paper out of an ensemble of size $M$ versus its rank $r$.
Figure~\ref{fig:zipf} shows the Zipf plot for some positions in the
astro-ph listing. We use the normalized rank $r/M$ instead of the rank
$r$ to account for the different lengths of astro-ph mailings. We bin
higher astro-ph positions to beat down the noise that comes from the
smaller number of articles at higher positions, i.e., from the smaller
number of astro-ph mailings that are longer than average.

The constant slope of the Zipf plots over a wide range of normalized
ranks confirms the power law nature of the citation distribution. One
clearly sees that the loci of the curves for positions 1 and 3 are
higher than those of e-prints further down the astro-ph listing with
an apparent continuous progression down for lower astro-ph positions.
The different loci of the curves correspond to the different
normalizations of the power law, i.e., the different number of
citations that e-prints at different positions in the astro-ph listing
get. We call this the \emph{positional citation effect} (PCE) and
adopt the following procedure to quantify the magnitude and
significance of the PCE.

We make a Zipf plot for the ensemble of all astro-ph e-prints and fit
a line $f(\ln(r/M)) = \beta \ln(r/M) + a$ to it in the range in which
the distribution follows a power law. In agreement with
\citet{1998EPJB....4..131R} we find $\beta = -0.48$ with a very small
error of $0.02$. Keeping $\beta$ fixed at this value we fit
$f(\ln(r/M))$ to the Zipf plots for each (binned) astro-ph position in
the range $-4.5 < \ln(r/M) < -2$. The value of $a$ is directly related
to the normalization of the power law $N(r/M) \propto
(r/M)^{\left(1+\frac{1}{\beta}\right)}$ and determines the loci of the
citation distributions in Fig.~\ref{fig:zipf}.

\begin{figure}
  \plotone{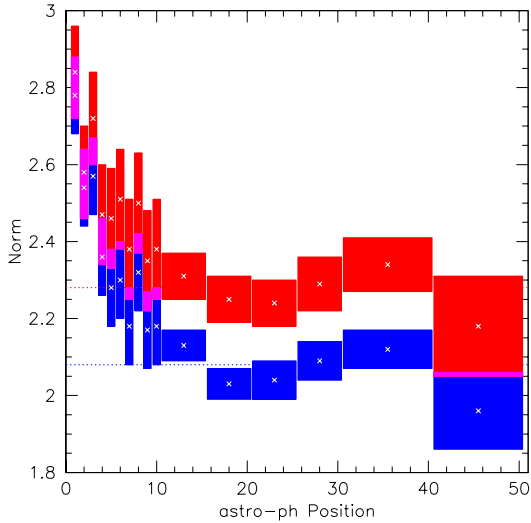}
  \caption{Normalization of the citation distribution power law. Blue
    boxes give the value of the power law normalization for all
    articles on astro-ph and its error. The widths of the boxes
    corresponds to the size of the bins. Red boxes show the same
    quantity but with the analysis restricted to the 7 core journals
    in Astronomy. The dotted horizontal lines are the normalization
    factors of the respective samples of all papers on astro-ph
    positions between $10$ and $40$.}
  \label{fig:zipf_coeff}
\end{figure}

Figure~\ref{fig:zipf_coeff} shows the results of this analysis for two
data sets. The blue boxes give the normalization for the ensemble of
all astro-ph e-prints, while the red boxes corresponds to those
e-prints that also appeared in one the core journals. Comparing the
normalization of the bins to that of the ensemble of e-prints on
astro-ph positions $10$--$40$, given by the dotted lines in
Fig.~\ref{fig:zipf_coeff}, we find that the PCE is significant for the
first six astro-ph positions for both data sets. The difference
between the normalization at position $1$ and the articles on
positions $10$--$40$ is significant at the $7\sigma$ level for all
e-prints and at the $4.7\sigma$ level for core journal articles.

We have now established that the PCE is present and highly
significant. Most scientists will, however, wonder how the different
normalizations of the Zipf law translate into citation counts. By
restricting an analysis of average citations to the range over which
the power law holds, we avoid both the tail of exceptionally highly
cited papers and the bulk of mostly ignored publications. We determine
average citation counts in the range $-4.5 < \ln(r/M) < -2$ by
integrating over the normalized Zipf distribution. We find an average
citation count of $95.4\pm11.4$ for core journal articles and of
$89.8\pm9.0$ for all e-prints on astro-ph position~$1$. Core journal
articles appearing at positions $10$--$40$ are on average cited
$54\pm1.6$ times, while the mean citation count for all e-prints at
these position is $44.6\pm0.9$. The overall higher number of citations
for articles from the core journals is in agreement with the findings
of \citet{2004BAAS...36.1654S}.

We propose three possible explanations for the observed dependence of
citation counts on astro-ph position:
\begin{itemize}
\item The Visibility Bias (VB) postulate -- Papers appearing at the
  top of the astro-ph listing are seen by more people and thus cited
  more often than those further down the list, where the attention of
  the astro-ph readers might decrease;
\item The Self-promotion Bias (SP) postulate -- Authors tend to
  promote their most important works, and thus most citable articles,
  by placing them at prominent positions;
\item The Geography Bias (GB) postulate -- The submission deadline
  preferrentially puts those authors at the top of the listing whose
  working hours coincide with the submission deadline. This group
  already has higher citation counts for other reasons.
\end{itemize}
The last postulate merits some further explanation.
\citet{2007EurRev..15..3S} noticed that US American authors have a
higher fraction of highly cited papers than their European colleagues.
The submission deadline for the arXiv e-print server is 16:00~EST/EDT.
This is within the normal working time of astronomers in all of the
USA, while it is outside working hours for European
astronomers.\footnote{In this we of course ignore the few individuals
  who do the traditional night time work of astronomers.}  Assuming
for the moment that GB is not caused by VB, i.e., that US Americans do
not get cited more because they are preferentially at the top of the
astro-ph listing, the PCE could be explained by the dominance of
American authors at the relevant moment in time. This explanation of
GB ignores astronomical communities outside Europe and the USA, but
they are comparably small.

To test the GB postulate we analyze the author affiliations of
e-prints appearing in core journals. We associate articles to one of
the following regions according to the country of the first
affiliation of the first author, where the number in parentheses
indicates the total number of articles from the respective region in
our database: Europe (6843), USA (5659), Asia (1693), North America
without USA (605), South America (412), Australia/Oceania (397).
Russia (166) and Turkey (32) are counted as Asian countries, Iceland
(3) as European. We treat the USA separately from the rest of North
America because \citet{2007EurRev..15..3S} made their observation only
for authors from the USA.

We repeat the analysis of the Zipf plots separately for authors from
Europe and from the USA. We find that the global power law indices for
both author sets are marginally inconsistent with each other. This
could potentially indicate a more fundamental difference between
papers written by European and American authors but our data set is
too small to draw firm conclusions and a more detailed study is beyond
the scope of this work. We settle for using the slopes of the Zipf
laws that were found for the respective sample.

\begin{figure}
  \plotone{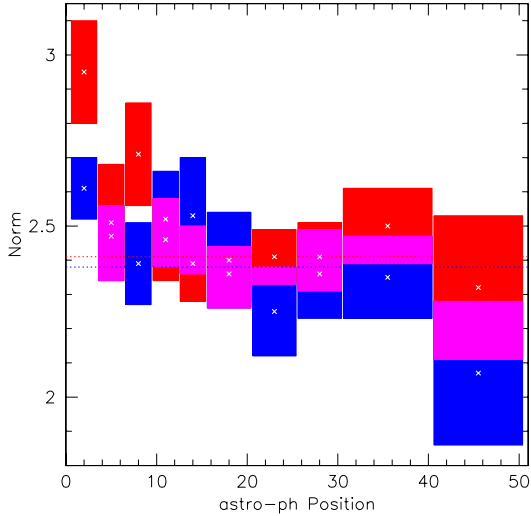}
  \caption{Normalization of the citation distribution power-law for
    European (red) and American (blue) authors of core journal papers.
    Note that the binning is coarser than in Fig.~\ref{fig:zipf_coeff}
    to account for the smaller number of articles in each sample.}
  \label{fig:zipf-pos-affiliation}
\end{figure}

The results of this Zipf plot analysis are shown in
Fig.~\ref{fig:zipf-pos-affiliation} and proof that the PCE is also
present for the sample of European authors. This rules out the GB
postulate as the single explanation for the observed PCE in the whole
sample. Note that the higher normalization of European authors does
not mean that these are more cited than their US American colleagues.
The difference is simply due to the different power law indices.

We emphasize that European authors who appear at the top of an
astro-ph listing had -- in all likelihood -- to submit their
manuscript well outside their normal working hours. It is thus
reasonable to assume that a conscious effort was made to gain this
position and to self-promote the work presented in these e-prints.
This is further supported by a comparison with the same statistics
made for US American authors. A significant increase of citations at
the top of the astro-ph listings is also seen for these authors but
the effect is not nearly as prominent as for their European
colleagues. A fair fraction of American authors will appear at or near
the top of the astro-ph listing just by chance, without any attempt at
self-promotion. They dilute any SP signal in this sample.

So far we could rule out GB as a significant contribution to the PCE
and found evidence for SP as a source of the PCE. Analyzing the
contribution of VB is much more difficult. In principle this would be
possible by examining the submission times of e-prints and grouping
them into two samples; one that is submitted so shortly after the
deadline that it is statistically expected to be self-promoted, and a
second one that is submitted long enough after the deadline to exclude
self-promotion. Unfortunately, the arXiv stores only the submission
time of the last replacement that was made without generating a new
version number. It does not keep the initial submission time of an
e-print. Without the latter it is not possible to disentangle VB and
SP.

Figure~\ref{fig:zipf_coeff} seems to show a marginally significant
drop of the power law normalization for articles at the bottom of very
long astro-ph mailings. This is most likely an artefact caused by the
growth of astro-ph over the period under investigation here. Far fewer
listings had more than 40 entries in 2002 and 2003 than in 2004 and
2005. These articles had less time to garner citations, an effect
which we, as mentioned previously, do not correct for. We find that
the decline in citation rates at the bottom vanishes if we restrict
the citation analysis to e-prints that appeared in 2004 and 2005.

\section{Summary and conclusions}
\label{sec:disc-concl}
We investigated the positional dependence of citation counts of
e-prints on their position in daily arXiv:astro-ph mailings. We found
that articles at or near the top of these listings receive
significantly more citations than articles at positions $10$ and
higher. This positional citation effect is present at the $7\sigma$
level for all e-prints and for the set of those e-prints also
appearing as articles in one of the core journals in Astronomy and
Astrophysics with a significance of $4.7\sigma$. Restricting the
analysis to the range over which the citation distribution follows a
power law, the difference in normalization factors translates to a
factor 2 difference in average citation counts. E-prints at the top of
the astro-ph listing are on average cited $89.8\pm9.0$ times while
articles between positions $10$ and $40$ receive only $44.6\pm0.9$
citations. The PCE is significant for the first six positions in the
astro-ph mailings.

We proposed three possible explanations for the observed PCE. By
analyzing the affiliations of authors we could exclude the hypothesis
that the PCE is caused by a geographical bias, namely by
preferentially putting US American authors at the top, who already
have higher citation rates than their European colleagues. Confirming
that the PCE is present and more pronounced for European authors, who
usually can gain the top position on astro-ph only by submitting
outside their normal working hours, we could conclude that
self-promotion plays a role in creating the PCE. 

We cannot present firm statistical evidence that the PCE at the top is
due to higher visibility. However, the PCE is present down to astro-ph
position 6, where the submission times of e-prints suggest that only a
very small fraction of manuscripts at this position was submitted with
the intention of self-promotion. Of course, this statement has to be
made with some care since the submission times reported by arXiv are
often not the original submission times as explained above. We,
however, observe that less than $5\%$ of e-prints at position 6 have
reported submission times within half an hour after the deadline,
while this fraction for e-prints at the top position is $62\%$,
suggesting that VB plays some role.

A situation in which the number of citations a publication gets does
not solely depend on its merits is unsatisfactory. One may be tempted
to suggest to randomize the order of e-prints in an astro-ph mailing,
instead of listing them sorted by submission time. This proposal has
severe weaknesses. On the one hand, the PCE in a randomized listing
would only vanish if the observed effect is completely due to SP bias,
since any visibility effects would immediately recur in a randomized
listing. On the other hand, if in fact SP bias is the sole cause of
the PCE there is no reason to change anything. It is important to
understand that SP means that papers at the top are on average
intrinsically more citable -- or better by some metric -- than papers
further down the list. A PCE caused only by SP bias is not in
contradiction to merit based citation counts. Hence, randomizing the
astro-ph listings would not solve the visibility problem and only
remedy a non-existing problem.

Any visibility problem is connected to the length of astro-ph
listings. Shorter listing would naturally give the same or almost the
same visibility to all articles, irrespective of their position.
Looking at the arXiv submission history, it is obvious that the number
of astro-ph e-prints will not decrease but only increase with time. A
possible method leading to shorter listings would be to split astro-ph
into several subject-classes, like the other three big archives
``hep'' (High-Energy Physics), ``cond-mat'' (Condensed Matter), and
``math'' (Mathematics) are subdivided into smaller categories.
Ideally, most researchers would have to look at only one or two more
specific and hence much shorter listings per day. For example, a
cosmologist could avoid seeing e-prints on Solar System objects if
he/she chooses not to read a possible Solar System subject-class.

A recent approach to tackle the increasing volume of astro-ph is the
arxivsorter\footnote{\texttt{http://arxivsorter.org/}}
\citep{2007Arxivsorter..M}. Arxivsorter aims to sort daily, recent, or
monthly astro-ph listings by relevance to an individual reader. The
underlying idea is that scientists through co-authorship form an
interconnected network of authors. By specifying a few authors
relevant to a reader's fields of interest, the ``proximity'' of a new
e-print in the author network can be calculated. This proximity seems
to be a good proxy for relevance to a reader's interests. Arxivsorter
has some problems with non-unique -- mostly Chinese -- names, and a
small percentage of authors not connected to the global cluster of
authors. It is worth pointing out that arxivsorter just re-orders the
papers without any loss of information, a clear advantage over a
possible split into subject-classes.

\acknowledgements This research has made use of NASA's Astrophysics
Data System Bibliographic Services. I am very grateful to Maryam
Modjaz, who identified a crucial error in a previous version of the
analysis presented here. Brice M\'enard and the referee Michael Kurtz
gave a number of excellent suggestions, which improved this paper. I
thank Uta Grothkopf for a careful reading of the manuscript.

\end{document}